\documentclass[%
aps,
amsmath,amssymb,
reprint,%
]{revtex4-1}

\usepackage{graphicx}
\usepackage{dcolumn}
\usepackage{bm}

\usepackage[utf8]{inputenc}
\usepackage[T1]{fontenc}
\usepackage{mathptmx}
\usepackage{color}
\usepackage{lmodern}

\usepackage{filecontents}

\begin{document}

\title{Thermoelectricity of near-resonant tunnel junctions and their near-Carnot efficiency}

\author{Matthias A. Popp}


\author{Andr\'e Erpenbeck}
\author{Heiko B. Weber}

\email[]{heiko.weber@fau.de}

\affiliation{Department Physik,
	Friedrich-Alexander-Universit\"at Erlangen-N\"urnberg, Staudtstr.\ 7,
	91058 Erlangen, Germany}

\homepage[http://www.lap.nat.fau.de]
\date{\today}

\begin{abstract}
	\textbf{The resonant tunneling model is the simplest model for describing electronic transport through nanoscale objects like individual molecules. A complete understanding includes not only charge transport but also thermal transport and their intricate interplay. Key linear response observables are the electrical conductance G and the Seebeck coefficient S. Here we present experiments on unspecified resonant tunnel junctions and molecular junctions that uncover correlations between $G$ and $S$, in particular rigid boundaries for $S(G)$. We find that these correlations can be consistently understood by the single-level resonant tunneling model, with excellent match to experiments. In this framework, measuring $I(V)$ and $S$ for a given junction provides access to the full thermoelectric characterization of the electronic system. A remarkable result is that without targeted chemical design, molecular junctions can expose thermoelectric conversion efficiencies which are close to the Carnot limit. This insight allows to provide design rules for optimized thermoelectric efficiency.}  
	
\end{abstract}

\maketitle

\section{Introduction}

Thermoelectric transport i.e. the unified consideration of heat and charge transport bears intrinsic correlations. The most famous among them is the Wiedemann-Franz law \cite{Franz1853}, which connects heat conductivity and electrical conductivity in linear response, not only in the ohmic regime but also in nanoscale metallic junctions \cite{Cui2017}. At functional interfaces with a sharp voltage and temperature drop ($V$,$\Delta T=T_H-T_C$) and nonlinear transmission function $\tau(E)$ however, it looses validity.
There, optimum thermoelectric power conversion of the electronic system is achieved for Dirac-delta like $\tau(E)$ \cite{Mahan1996}. This picture has been refined: in the presence of finite heat conductance by other channels (e.g. phonons, photons), boxcar functions promise maximum efficiency at finite power \cite{Whitney2014}.

Here, we focus on the resonant tunneling regime \cite{Cuevas2010}, which provides a good description for charge transport  through quantum dots and, in particular, molecular junctions.
A recent summary on thermoelectricity in molecular junctions is given in \cite{Rincon-Garcia2016,Cui2017a}.  There, the junction conductance $G$ ranged from $10^{-5}..\; 10^0\;G_0$ (with the conductance quantum $G_0$) and corresponding Seebeck coefficients $S=- V_{th} / \Delta T$ (also termed thermopower, with thermovoltage $V_{th}$) are typically below 30 $\mu\mathrm{V}/\mathrm{K}$. This is, as it will turn out in this paper, a rather low value. When in addition electrostatic gates are present, parameter sets could be tuned to provide higher $S$ \cite{Gehring2017,Kim2014}. We present experiments on resonant tunnel junctions / molecular junctions and theory that relate $G$ and $S$ and address the question whether correlations and/or design rules can be recognized.

The recent development of squeezable nanojunctions (SNJ)\cite{Popp2019} gives access to such phenomena in the resonant tunneling regime with either unspecified (contamination induced) junctions, metallic nanoparticle junctions and/or molecular tunnel junctions,  in a broad range ($10^{-4}..\; 10^0\;G_0$). With this technique $S$ and $G$ can immediately be compared. These two quantities can be considered as one datapair on the very same atomistic structure because of the outstanding stability of the junctions. Hence, correlations between  $G$ and $S$ can be extracted.

\begin{figure}[h]
	
	\centering
	\includegraphics[width=\columnwidth]{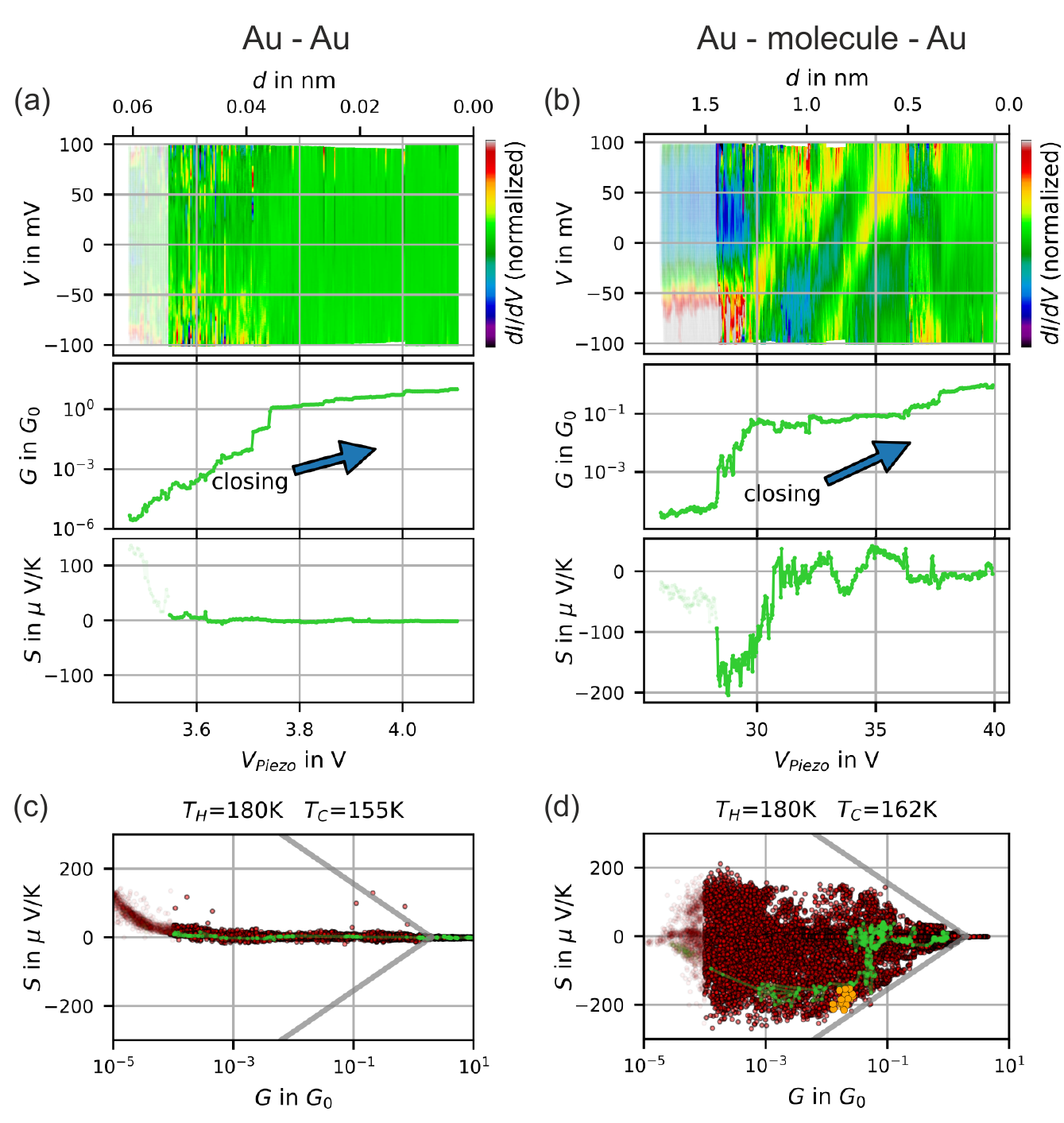}
	\caption{\textbf{Electric and thermoelectric characterization.} (a) and (b) show $\mathrm{d}I/\mathrm{d}V$ curves (color coded), conductances $G$ and Seebeck coefficients $S$ measured during selected closing cycles for (a) a bare gold-gold junction and (b) a junction with molecules (see SI) applied to the electrodes. $\mathrm{d}I/\mathrm{d}V$ values were normalized for better visibility of the curve shape. (c) and (d) show correlations of $S$ and $G$ using the data of the whole ensembles enclosing 21 (27) opening and closing cycles. Green dots correspond to the data in (a), (b). Orange dots mark a sub-ensemble with high conversion efficiency. Gray lines are eye guides which separate $G-S$ pairs from exclusion areas. \label{FIG2}}
	
\end{figure}

\section{Experimental Observations}

We measure individual $I(V)$ curves during stepwise opening and closing cycles. $\mathrm{d}I/\mathrm{d}V$ curves underlying one closing cycle of a metallic junction are depicted in FIG. \ref{FIG2} (a). From these data $G$ can be derived, yielding closing curves similar to previous nanojunction experiments. In addition, $S$ is recorded stepwise such that each $G-S$ pair can immediately be related. An individual dataset consists of $I(V)$ and $S$. About 20000 of such datasets (20-30 opening/closing cycles) are considered as an ensemble. FIG. \ref{FIG2} (c) shows the full ensemble of $G-S$ datapairs, each of which is represented as a red dot.

The left and right column in FIG. \ref{FIG2}. show data recorded in two independent experiments, representing the full range of phenomena that we observe. We first describe pure gold samples under high vacuum conditions (FIG.\ref{FIG2}. (a),(c)). $\mathrm{d}I/\mathrm{d}V$ curves are pretty much constant, the Seebeck coefficient for $G>G_0$ i.e. in the metallic regime is below 10 $\frac{\mu \mathrm{V}}{\mathrm{K}}$, the overall observations are in full agreement with previous measurements including step wise reduction of conductance, pronounced plateaus at $G=G_0$ \cite{Cuevas2010,Krans1995} and Seebeck coefficients as described by \cite{Ludoph1999,Evangeli2015}. Below $G_0$ i.e. in the tunneling regime we find slightly enhanced thermopowers that scatter around $+1\;\frac{\mu\mathrm{V}}{\mathrm{K}}$ with a standard deviation of  $6\;\frac{\mu\mathrm{V}}{\mathrm{K}}$. Below $10^{-4}\;G_0$ deviations due to experimental artifacts were observed, more precisely a voltage that results from finite offset-currents of the voltage amplifier, which cannot fully be compensated for. From the measurement we can derive that it is 1 pA for our setup.

The right column in FIG. \ref{FIG2} shows data obtained with intentionally added molecules (also FIG. S1 (a)). FIG.1 (b) shows data of a selected closing curve with resonances shifting through the fermi level. Direct evidence is given by peaks in $\mathrm{d}I/\mathrm{d}V$ at zero Voltage \cite{Popp2019}. Additionally, closing curves of both G and S are shown. Note that highly structured $\mathrm{d}I/\mathrm{d}V$ coincide with high values for $S$. The corresponding $G-S$ data of the full ensemble are displayed in (d). For this (near) resonant tunneling case, a qualitatively different picture is found: 
the Seebeck coefficient covers a wide span of values up to and even beyond 200 $\mu \mathrm{V/K}$.  This is more than an order of magnitude larger than in the (flat) tunneling regime. It should be mentioned that no further selection or filtering was applied. The data explore large areas of the $G-S$ manifold like if there was no correlation. However, there are clear exclusion areas, the boundaries of which are indicated by gray solid lines as guide to the eyes. They are kept identical in all $G-S$ plots throughout this manuscript. We emphasize two unexpected findings uncovered by this plot: first, the Seebeck coefficients in the resonant tunneling regime are significantly higher than reported in earlier ungated single-molecule studies \cite{Rincon-Garcia2016,Cui2017a}. Second, $S$ is limited by boundaries. As it turned out later, very similar findings could be obtained by nanoparticle \cite{Schmutzler2019} junctions and unspecified contamination states (FIG. S1 (b),(d)).

FIG.\ref{FIG2}. (c) and (d) represent two limiting cases of our observed ensembles: whereas in FIG.\ref{FIG2}.  (c) (pure gold) the data accumulate close to the $S=0$ axis, the data in (f) (with molecules  providing resonant tunneling behavior) show a broader spread of $S$ values. Further experiments delivered ensembles that were bouncing in between both cases or did not fill the full accessible part of the $G-S$ manifold, presumably because of incomplete sampling within the finite measurement time. More data can be seen in SI. While we focus on the ensembles, a plethora of individual $I(V)$ characteristics is not considered here in detail, but are available as raw data in \cite{Popp}

\section{Theoretical Analysis}

\begin{figure}
	
	\centering
	\includegraphics[width=\columnwidth]{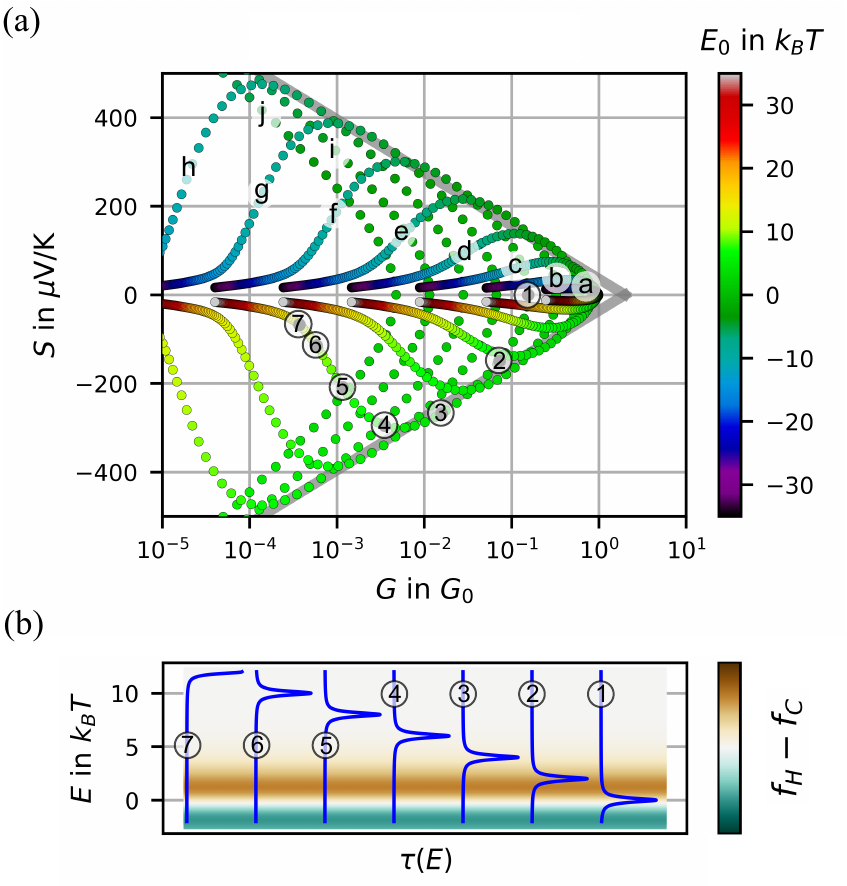}
	\caption{\textbf{Conductance and Seebeck coefficient in the resonant level model.} (a) Seebeck coefficients and conductances calculated within the resonant level model. The trajectories a-j  are curves of constant $\Gamma$, varied logarithmically from $10\,k_BT $ to $3\cdot 10^{-3}\,k_BT$. Within each trajectory the position of the energy level $E_0$ was varied in equidistant steps indicated by the color scale. The gray lines are the very same eye guides as in FIG.\ref{FIG2}. which mark the boundaries to the excluded area. (b) Transmission functions $\tau(E)$ along the trajectory with $\Gamma=1.1\cdot 10^{-1}\; k_BT$. The labels 1-7 correspond to the labels in (a) \label{FIG3} }
	
\end{figure}

The experimental findings are discussed in the framework of the resonant tunneling model within the Landauer transport picture \cite{Cuevas2010,Rincon-Garcia2016,Cui2017}. We choose a Lorentz-shaped transmission function
\begin{equation}
\tau(E)=\frac{4\Gamma^2}{(E-E_0)^2+4\Gamma^2}
\end{equation}
at energy $E_0$ and width $\Gamma$. The latter reflects broadening due to coupling to the leads, which we assume to be symmetric for simplicity. In this picture $G$ and $S$ are calculated as 
\begin{equation}
G = \frac{2e^2}{h}\int-\frac{\partial f}{\partial E}\left(\frac{T_H+T_C}{2},E\right)\tau(E)dE,
\label{eq:conductivity}
\end{equation}
\begin{equation}
S=\frac{1}{G} \frac{-2e}{h}\int [f(T_H,E)-f(T_C,E)] \tau(E)dE. \label{eq: S}
\end{equation}

We sample the parameter space by varying $E_0$ and $\Gamma$ for fixed $T_H$ and $T_C$. We choose a representation of $S$ vs. $\log(G)$, motivated by the experiment. The result is shown in FIG.\ref{FIG3}. where we choose ten different trajectories of constant $\Gamma$ (logarithmically equidistant) and vary $E_0$, the latter can be recognized by its color code. The trajectories have a rounded arrow-head shape. We start the discussion with the case $E_0=E_F=0$ (resonant case, \textcircled{1} in FIG.\ref{FIG3}.(b)), at which the fully symmetric Fermi distribution of electrons along with the symmetric transmission function of the level results in zero thermopower. When however lifting $E_0$ above the Fermi-energy, such that  the Lorentz resonance is still within the thermal window (near-resonant case \textcircled{2}, \textcircled{3}), the Seebeck coefficient rises with $E_0$ and reaches remarkably high values, fully compatible with the experimental values. Beyond position \textcircled{4}, when the Lorentz resonance moves far out of the thermal window  $S$ decreases again (off-resonant case \textcircled{5}-\textcircled{7}). In this limit the quadratic roll-off of the Lorentzian gives the dominant contribution in Eq. (\ref{eq: S}). This overall behavior is fully symmetric for positive and negative $E_0$. Sampling the full $E_0,\; \Gamma$ parameter space, the exclusion areas observed in experiments are recovered without any further assumptions. The boundaries are formed by two enveloping lines which are identical with the gray lines in FIG.\ref{FIG2}. A closer look reveals, however, that the envelope is slightly curved, the gray straight line is therefore rather a guide to the eye. Notably any $G-S$ data pair between the envelopes can be generated. This finding is robust when the assumption of symmetric coupling to the left and right electrode is lifted: then the overall pattern is shifted towards lower conductances (the exclusion area is maintained). Similarly when in a simple gedanken experiment two near-resonant tunneling paths in parallel form the junction, $S$ is the same as for a single one, but the conductance $G$ is doubled. This would then enter into the excluded area, which would, however, be barely visible on the logarithmic scale. We conclude that, despite its simplicity, the single resonant level model is suited to describe the experimental correlations between $S$ and $G$. 

The shape of the chosen trajectories in FIG.\ref{FIG3}. is not purely artificial.As an anecdotic example we highlighted one trajectory in FIG.\ref{FIG2}(d) in green color, which corresponds to the closing curves in (b). In the range of $V_{Piezo} = 28.. \;31$ V the energetic position of the resonance moves towards the Fermi-Energy accompanied by $G$ increasing from $10^{-3}.. \,10^{-1}\;\; G_0$ and S varying between $-200 .. \;0\;\; \mu\mathrm{V}/\mathrm{K}$. This may be explained by the continuous evolution of the all-important local environment of the molecule (electrostatic, strain, shape \cite{Perrin2013,Kim2014}).In any case high Seebeck coefficients coincide with near-resonant features in the $\mathrm{d}I/\mathrm{d}V$.

\section{Implications}
Having a suitable model at hand that describes essential aspects of thermoelectrical transport through near-resonant tunneling states, we further study properties and implications of this model. The Seebeck coefficient is often considered to be proportional to temperature $T$ (Mott formula). In the model under investigation this is recovered only in the off-resonant case. When however the resonance is closer to the Fermi level (near-resonant case), where $\tau(E)$ is strongly curved, the Mott assumptions are not fulfilled. $S$ can then be strongly nonlinear in $T$ (it remains, however, a linear response to $\Delta T$). Further the Wiedemann-Franz law can be rediscovered in selected areas of the $G-S$ pattern: It is valid in the off-resonant case (position \textcircled{7} and beyond), where both $\tau(E)$ and $\partial \tau(E)/\partial E$ are essentially constant at $E=E_F$. Next to the envelope \textcircled{3}, the Wiedemann-Franz ratio delivers the Lorentz number (see SI). 

Notably, the $G-S$ envelope remains untouched by variations of $T$. This can be understood by regarding the scaling behavior of the three energy scales $k_BT$, $E_0$, $\Gamma$. Upon multiplication of all three quantities with a common constant factor, the resulting $G$ and $S$ values are untouched, as well as the $G-S$ pattern, underscoring the generality of the concept. 

\onecolumngrid

\begin{figure}
	\centering
	\includegraphics[width=\columnwidth]{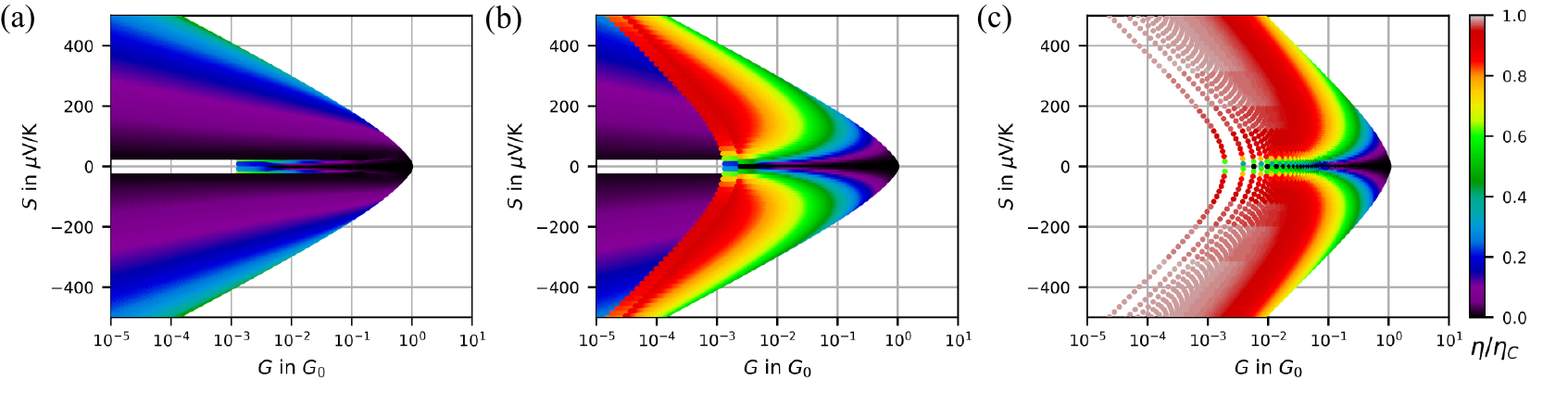}
	\caption{\textbf{Heat conversion efficiency of the electronic system.} Heat conversion efficiency $\eta$ of the electronic system calculated with respect to Carnot-efficiency $\eta_C$. (a) and (b) are calculated within the resonant level model. In (a) the off-resonant regime is in the foreground whereas in (b) the near-resonant regime is visible. (c) rectangular transmission function for comparison. Note that here also the excluded area differs. \label{FIG4}}
	
\end{figure}

\twocolumngrid
Next to large Seebeck coefficients and their boundaries, the heat conversion efficiency is of fundamental interest. It quantifies how much electrical power $P_{el}$ can be generated with respect to the invested heat flux $\dot{Q}_{in}$ necessary to maintain $T_H$:
\begin{equation}
\eta (V)= \frac{P_{el}(V)}{\dot{Q}_{in}(V)}=\frac{-V\cdot I(V)}{\dot{Q}_{in}(V)}
\end{equation}
in the bounds $0<V<V_{th}$
with 
\begin{equation}
I(V)= \frac{-2e}{h}\int[f_H(E+\frac{eV}{2})-f_C(E-\frac{eV}{2})] \tau(E)dE,
\label{eq:IV}
\end{equation} 
\begin{equation}
\dot{Q}_{in}(V)= 
\frac{2}{h}\int(E+\frac{eV}{2})[f_H(E+\frac{eV}{2})-f_C(E-\frac{eV}{2})] \tau(E)dE.
\label{eq:QV}
\end{equation}

Here, the voltage is assumed to drop symmetrically across the junction i.e. the electrochemical potential of the hot lead is assumed to be lowered by $\frac{eV}{2}$ and vice versa for the cold lead. As $\dot{Q}$ is not available in our experiment, we choose to calculate it from the very same transmission function. A similar procedure has been used by \cite {Josefsson2018,Josefsson2019} for quantum dots (energy scales 100 times smaller).

The maximum efficiency for given $\tau(E)$ is then calculated by numerically maximizing $\eta$ with respect to $V$. We normalize the efficiencies with respect to the Carnot efficiency $\eta_C=\frac{T_H-T_C}{T_H}$ that is the highest possible efficiency in accordance with the second law of thermodynamics. FIG.\ref{FIG4} (a) and (b) show color-coded numerical results, separately plotted for parameter regions corresponding to the off-resonant and the near-resonant case, respectively. It becomes immediately visible that efficiencies close to the Carnot limit (red color) are only possible in the near-resonant case. There the transmission is dominated by a delta-peak like $\tau(E)$, meaning that the integrands in equations (\ref{eq:IV}) and (\ref{eq:QV}) are evaluated only at $E_0$ \cite{Mahan1996,Whitney2014}. The envelope has again special features: here, the efficiency is $0.5 \;\eta_C$ when being sufficiently distant from the arrow head tip. Such astonishingly high efficiencies were reached by our experiments.Notably  these values  appeared without targeted design of the molecule, and even in unspecified contamination states. 
For a better classification, we compare the $G-S$ plot for resonant junctions with boxcar transmission profiles that are the profiles that are expected to be best in efficiencies \cite{Whitney2014}, cf. FIG.\ref{FIG4} (c). Indeed high efficiencies can be found at significantly higher conductance values which is favorable for thermoelectric heat conversion at high power output. Remarkable is, though the very similar over all pattern, even if the boundaries are shifted outwards. This underscores that the sharper the transmission function is, the better the efficiencies can be expected.

However, further contributions should be considered that reduce this efficiency: (i) an electronic contribution with flat $\tau(E)$ \cite{Popp2019} that provides an offset conduction $G_{off}$, enclosing unspecified spectrally broad  tunneling channels (electronic background transparency) and (ii) a vibrational heat conductance $G_{th,vib}$ in addition to the electronic heat conduction. Both do not contribute to thermovoltage. The electronic terms are accessible by analyzing the full data set including $I(V)$ out of which we can determine $E_0$, $\Gamma$, $\alpha$ and $G_{off}$ (see SI, \cite{Popp2019,Newville2014}). Evaluating a subensemble (orange dots in FIG.\ref{FIG2}. (d)), we find $\eta_{res}=0.5\;\eta_C$ for the resonance-only case which leads to a thermoelectric figure of merit $ZT_{res}\approx 8$. When $G_{off}$ is included, it reduces to $\eta_{el}=0.3\;\eta_C$, ($ZT_{el}\approx 2.5$). We have no value for this particular molecule, but introduce the only experimentally determined value of vibrational heat conductance (measured with alkane molecules):  $G_{th,vib}\approx 20\frac{\mathrm{pW}}{\mathrm{K}}$ \cite{Cui2019,Klockner2017} which finally leads to  $\eta_{realistic}=0.1\;\eta_C$ ($ZT_{realistic}\approx 0.5$).
These values are competitive compared to recent $ZT$ record values \cite{Venkatasubramanian2001,Zhao2014,Tan2016,Hinterleitner2019}.
\begin{figure}
	\centering
	\includegraphics[width=\columnwidth]{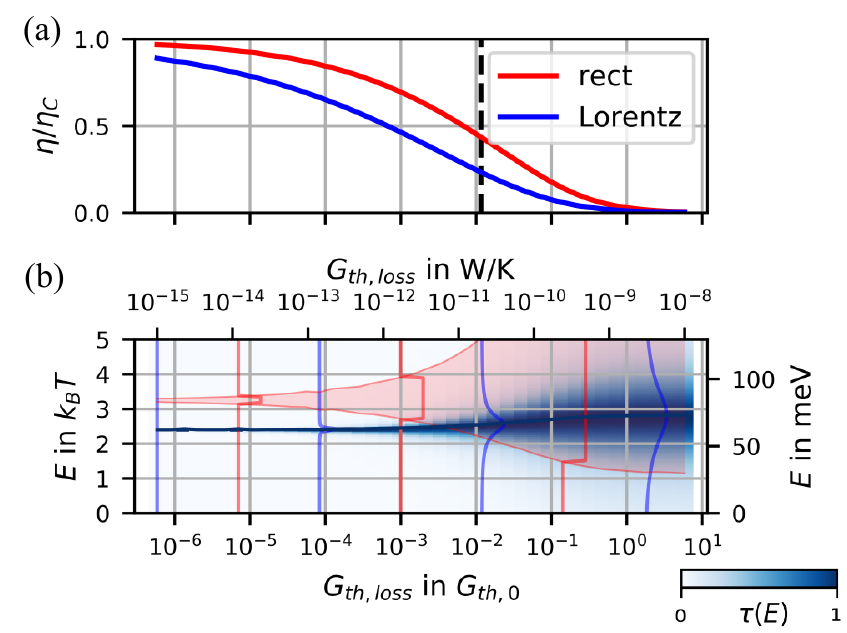}
	\caption{\textbf{Efficiencies with optimized transmission functions.} (a)  Optimal values of the thermoelectric conversion efficiency for rectangular and Lorentz shaped transmission functions, plotted as a function of $G_{th,loss}$. (b) The underlying optimized transmission functions.This plot is temperature invariant when choosing units of $k_B T$ and the (electrical) thermal conductance quantum \cite{Cui2017b} $G_{th,0}=\frac{2\pi^2k_B^2T}{3h}$ (left and lower scale). The right and upper scale denominate values at room temperature (T=300K). \label{FIG5}}
	
\end{figure}

On the first sight, an optimization of efficiency would reduce $G_{th,vib}$ and keep the electronic system untouched. Our model, however, gives access to further potential optimization: for a given $G_{th,loss}$ (including $G_{th,vib} $ and the thermal loss due to $G_{off}$), we search for parameters  $E_0,\;\Gamma$ such that the efficiency is optimized. The numerical results are displayed in FIG.\ref{FIG5}. In (a), the optimized $\eta$ is plotted both for the resonant tunneling model (blue) and a rectangular transmission window as a function of $G_{th,loss}$ in a broad range. 
Along with the results for the resonant level model we evaluated a hypothetical system with boxcar-type $\tau(E)$ which has shown to be the theoretically most efficient transmission function.

Hence, improvements targeting high efficiency interfaces should first ensure near-resonant electronic conditions, more explicitly $E_0\approx 2.5\;k_BT$ and $\Gamma\approx 0.5\;k_BT$. Favorable is, of course, the design of low heat conductance. In case $G_{th,loss}$ drops below $10^{-9} \frac{\mathrm{W}}{\mathrm{K}}$ (at 300 K), $\Gamma$ should be matched accordingly. Due to our numerical analysis this complicated design challenge becomes now better tractable: we lay down optimized parameters in a look-up table, see table S2.

\section{Conclusions}

We present an experiment that is able to immediately compare measured Seebeck coefficients $S$  with the conductance $G$ of resonant tunnel junctions, including single-molecule junctions. Having sampled a large parameter space of the resonant tunnel model both in theory and experiment we have explored correlations between $G$ and $S$. It is demonstrated that $I(V)$ and Seebeck coefficients can be invoked to determine thermoelectric efficiency. We provide design rules for optimized efficiencies which are indeed remarkably close to the Carnot limit. 

\section{Methods}

\begin{figure}
	
	\centering
	\includegraphics[width=0.8\columnwidth]{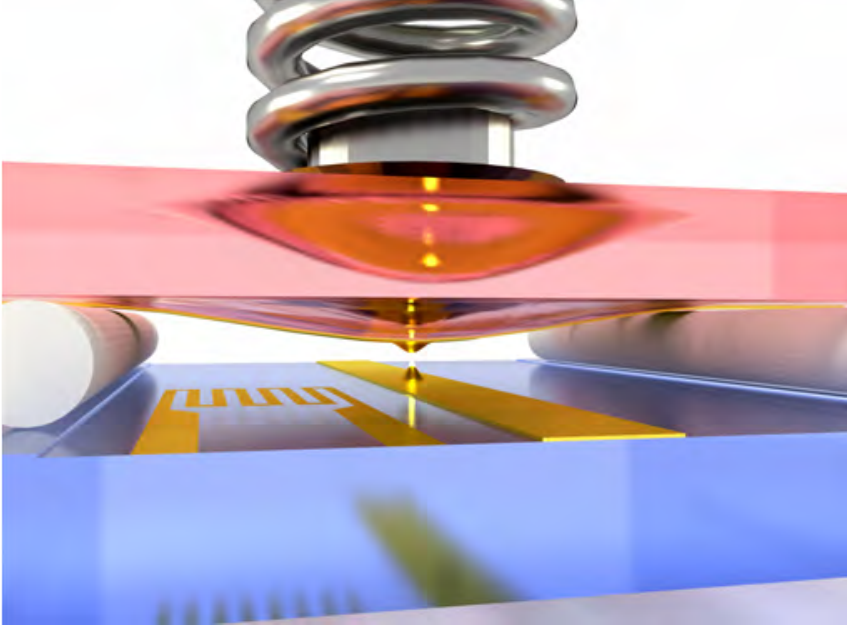}
	\caption{\textbf{Scheme of the squeezable nanojunction (SNJ).} Two SiC chips with electrodes are placed in a sandwich configuration. The ultra-stable distance between the electrodes is adjusted via a piezo-spring mechanism. Chip temperatures $T_H$ and $T_C$ are monitored via on-chip resistance thermometry. \label{FIG1}}
	
\end{figure}

The SNJ consists of two silicon carbide (SiC) chips in a sandwich configuration see FIG.\ref{FIG1}. By a piezo/spring mechanism they are compressed such that the distance at the touching point is controlled with extremely high stability and resolution (\cite{Popp2019}). The excellent heat conductivity of SiC guarantees that both entire chips have a defined temperature, which is explicitly measured and contributes $\Delta T$. By adding two lithographically defined gold electrodes also the electrical potentials can be controlled. Hence, each electrode is a thermal as well as electrical reservoir, between which nanojunctions can be established.
The electrical measurement is described in \cite{Popp2019}. $G$ is determined from a linear fit to the $I(V)$ in the voltage window -100... 100 mV. $S$ is derived by measuring the thermovoltage $V_{th}$ and an explicit measurement of $T_H$ and $T_C$ on the two SiC chips.
Data acquisition in FIG.\ref{FIG2}.(d) follows the same protocol as in (c) but this time molecules were dropcasted on the electrodes prior to cooling down. These data occurred after "activating" the junction which means that several I(V) sweeps in a voltage window of $\pm800$ mV were performed. The conductance measurement was performed again in the -100.. 100 mV window.

\section*{Supplemental Information}

See Supplemental Information for extended data, calculations on the Wiedemann-Franz law, description of fit routine, a look-up table as a different representation of FIG. \ref{FIG5}
\\
The full raw data underlying all figures in the main manuscript and SI are availible under \cite{Popp}.

\section*{Acknowledgments}

This work was supported by the Deutsche Forschungsgemeinschaft (DFG), Projektnummer 182849149 (SFB 953).
We thank Agustin Molina-Ontoria and Nazario Martin for synthesis of the molecules and Tobias Zech, member of the RTG1986, Erlangen for preparation of Au-nanoparticles. Further we thank Fabian Pauly for useful discussions.

\section*{Author contributions}

M.A.P. and H.B.W. conceived the experiment. M.A.P. performed the experiment and developed the theoretical framework, complemented by calculations from A.E.. All authors discussed the results scientifically, M.A.P. and H.B.W. wrote the manuscript with input from A.E..

\section*{References}
\bibliography{library}

\end{document}